\documentclass[letterpaper]{article} 
\usepackage{aaai2026}  
\usepackage{times}  
\usepackage{helvet}  
\usepackage{courier}  
\usepackage[hyphens]{url}  
\usepackage{graphicx} 
\usepackage{natbib}  
\usepackage{caption} 
\usepackage{algorithm}
\usepackage{algorithmic}

\usepackage{newfloat}
\usepackage{listings}
\DeclareCaptionStyle{ruled}{labelfont=normalfont,labelsep=colon,strut=off} 
\floatstyle{ruled}
\newfloat{listing}{tb}{lst}{}
\floatname{listing}{Listing}
\usepackage[colorinlistoftodos]{todonotes}

\usepackage{xspace}

\newcommand{\nop}[1]{}

\newcommand{\system}{\textsc{Thucy}\xspace}

\title{Thucy: An LLM-based Multi-Agent System for Claim Verification across Relational Databases}

\author{
    Michael Theologitis\textsuperscript{\rm 1},
    Dan Suciu\textsuperscript{\rm 1}
}
\affiliations{
    
    \textsuperscript{\rm 1}University of Washington \\
    Paul G. Allen School of Computer Science \& Engineering \\
    \{mthe, suciu\}@cs.washington.edu
}

\usepackage{bibentry}

\usepackage{color}
\usepackage{booktabs}

\usepackage[cachedir=.]{minted}
\usepackage{arydshln} 
\usepackage{pifont} 
\usepackage{lipsum} 

\newcommand\eat[1]{}
\usepackage{afterpage}
\usepackage{tikz}  
\newcommand*\circled[1]{\tikz[baseline=(char.base)]{
            \node[shape=circle,draw,inner sep=0.5pt] (char) {#1};}} 
\definecolor{greenish}{RGB}{102,204,0}
\definecolor{brownish}{RGB}{204,102,0}
\definecolor{redish}{RGB}{204,0,0}
\definecolor{yamlgreen}{RGB}{0.0, 153, 0.0} 
\usepackage{textcomp} 

\begin{document}

\maketitle

\begin{abstract}

In today's age, it is becoming increasingly difficult to decipher truth from lies. Every day, politicians, media outlets, and public figures make conflicting claims---often about topics that can, in principle, be verified against structured data. For instance, statements about crime rates, economic growth or healthcare can all be verified against official public records and structured datasets. Building a system that can automatically do that would have sounded like science fiction just a few years ago. Yet, with the extraordinary progress in LLMs and agentic AI, this is now within reach. Still, there remains a striking gap between what is technically possible and what is being demonstrated by recent work. Most existing verification systems operate only on small, single-table databases---typically a few hundred rows---that conveniently fit within an LLM's context window. 

In this paper we report our progress on \system, the first cross-database, cross-table multi-agent claim verification system that also provides concrete evidence for each verification verdict. \system remains completely agnostic to the underlying data sources before deployment and must therefore autonomously discover, inspect, and reason over all available relational databases to verify claims. Importantly, \system also reports the exact SQL queries that support its verdict (whether the claim is accurate or not) offering full transparency to expert users familiar with SQL. When evaluated on the TabFact dataset---the standard benchmark for fact verification over structured data---\system surpasses the previous state of the art by 5.6 percentage points in accuracy (94.3\% vs. 88.7\%).

\end{abstract}

 \begin{links}
     \link{Code}{https://github.com/michaeltheologitis/thucy}
 \end{links}

\section{Introduction}\label{sec:intro}


In the Annual Report released last year by the~\citet{SeattleCityAnnualReport_2024}, we read the following:
\begin{quote}
\textit{I am pleased to acknowledge that 2024 saw a reduction in property crime and violent crime in Seattle. \\ --- Ann Davison, City Attorney}
\end{quote}
However, for many residents of Seattle, this statement might not quite match their lived experience. The natural instinct is to want to find out more. Was crime really down in 2024? And if so, by how much---and according to which source?

It turns out, the City of Seattle publicly provides an official crime dataset~\cite{SeattleCrimeDataset}---with all crimes from 2008 until now---that is structured, detailed, and updated almost daily. In principle, that is all you would need to verify such a claim. In practice, though, very few people ever try. Most will simply take the statement at face value and move on, keeping the comforting thought that ``Seattle is safer now'' somewhere in the back of their mind.

A few more curious and determined souls might go a step further, dig around, discover the dataset, and even download it. Then reality hits: it is technical, messy, and not exactly friendly to non-specialists. So they, too, eventually give up. And so the claim remains---unchecked, unchallenged, and protected by the technical complexity of verification.

In this work, we present a multi-agent system called \system that takes over the verification process once the user has obtained the structured data and imported it into a relational database. From that point on, \system figures everything out: it autonomously explores the available data sources, reasoning over them on the fly to produce a verdict and supporting evidence. 

In our example, we can simply download the City of Seattle's official crime dataset, load it into a SQL database, and invoke \system with the verbatim claim of Ann Davison for verification. \system takes care of the rest---no further clarifications are needed. In fact, \system is completely agnostic to the underlying data environment before deployment.


We draw inspiration from the work of Thucy(dides), the Athenian historian (460--400 BC) who wrote the \textit{History of the Peloponnesian War} between Sparta and Athens. ``Thucydides has been dubbed the father of \textit{scientific history} by those who accept his claims to have applied strict standards of impartiality and evidence-gathering''~\cite{wiki:Thycidides}.

Following Thucydides' example of reporting, \system's job is twofold: \circled{1} provide a verification verdict (whether the claim is supported or not based on the available data), and \circled{2} return a report together with  SQL queries that explain its findings. By returning the explanations in the form of SQL queries, \system empowers the data analyst to modify these SQL queries and explore the claim further.  For example they can ``roll-up'' by checking if crime of all types has decreased in Seattle in 2024 (not just property and violent crime), or to ``drill down'' and check how crime changed in 2024 for each Seattle neighborhood.

\begin{figure*}[htbp]
  \centering
  \includegraphics[scale=0.8]{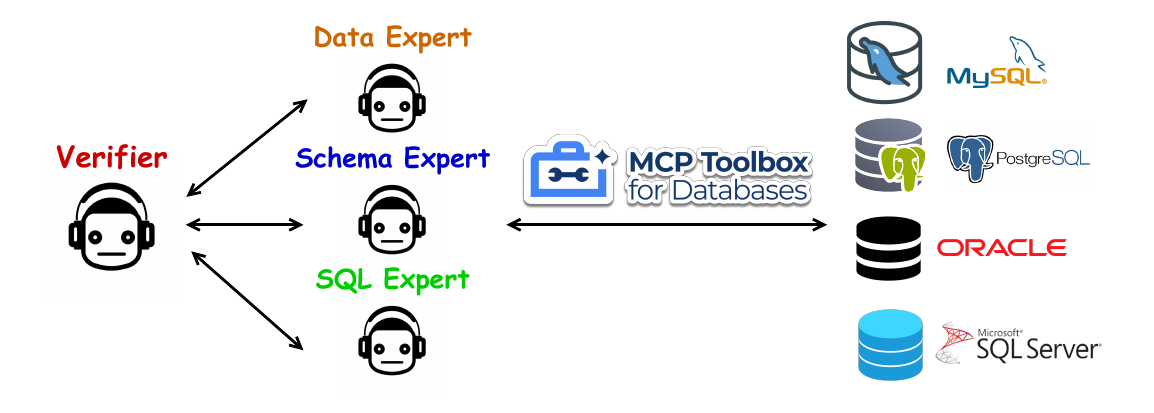}
  \caption{The architecture of \system, a multi-agent system led by the \textcolor{redish}{\textit{Verifier}}. Its job is to verify NL claims grounded in relational databases and report the corresponding SQL evidence. The \textcolor{redish}{\textit{Verifier}} coordinates three expert agents: the \textcolor{brownish}{\textit{Data Expert}}, which summarizes available data sources; the \textcolor{blue}{\textit{Schema Expert}}, which answers schema-related questions; and the \textcolor{greenish}{\textit{SQL Expert}}, which writes and executes SQL queries to obtain verifiable answers. The data layer follows a plug-and-play design and can include any number of relational databases---each potentially containing many tables---with PostgreSQL, MySQL, SQL Server, and Oracle shown here only as examples. \system remains fully agnostic to the underlying data sources. The agents must therefore operate in an open-ended environment, discovering and reasoning about available data as they encounter it. The experts interact with these relational databases through specialized tools managed via Google's MCP Toolbox. Adding or removing databases is straightforward: it simple involves adding or removing the corresponding tool from the toolbox.}
  \label{fig:architecture}
\end{figure*}

\section{Architecture}\label{sec:architecture}

In this section, we describe the architecture of \system. We start by discussing the data sources and our minimal assumptions about them. Then, we go over the recent standardized ways modern AI agents connect to databases. Finally, we delve into the details of our multi-agent system (Figure~\ref{fig:architecture}). 

Throughout this section, we aim to be as informative as possible about the unique ways multi-agent systems must navigate relational databases. Table~\ref{tab:tabfact_capabilities} summarizes, at a high level, how \system differs from prior systems that operate over structured data. Beyond explaining how our system works, our goal is to also make clear the rationale behind the design choices that made \system possible. Doing so naturally requires unpacking some of the subtleties of relational databases along the way.

\begin{table*}[htbp]
\centering
\caption{Capabilities of different LLM-based systems for \textit{fact verification} over structured data. Cross-Table and Cross-Database refer to a system's ability to verify claims that span multiple tables or databases. \textit{Interpretable} means that users can understand the reasoning behind the model's verdict. \textit{Verifiable} goes a step further---it allows users to reproduce the verification process (e.g., providing the exact Python or SQL commands), eliminating any suspicion of hallucinations. Finally, \textit{Source-Agnostic} indicates that the system can operate without prior knowledge of its data environment, figuring out everything from scratch.}
\begin{tabular}{l c c c c c}
\toprule
\textbf{Method} & \textbf{Cross-Table}& \textbf{Cross-Database} & \textbf{Interpretable} & \textbf{Verifiable} & \textbf{Source-Agnostic} \\
\midrule
BINDER~\cite{DBLP:conf/iclr/ChengX0LNHXROZS23} & \ding{55} & \ding{55} & \ding{51} & \ding{55} & \ding{55} \\
DATER~\cite{DBLP:conf/sigir/YeHYLHL23} & \ding{55} & \ding{55} &  \ding{51} & \ding{55} & \ding{55} \\
CoTable~\cite{DBLP:conf/iclr/0002ZLEP0MFSLP24} &  \ding{55} & \ding{55} &  \ding{51} & \ding{55} & \ding{55} \\
ReAcTable~\cite{DBLP:journals/pvldb/ZhangHFCDP24} & \ding{55} & \ding{55} &  \ding{55} & \ding{55} & \ding{55} \\
AutoTQA~\cite{DBLP:journals/pvldb/ZhuCXLSZSTL24} & \ding{51} & \ding{51} &  \ding{55} & \ding{55} & \ding{55} \\
POS~\cite{DBLP:journals/tmlr/NguyenBSKNL25} & \ding{55} & \ding{55} &  \ding{51} & \ding{55} & \ding{55} \\
\midrule
\system (ours) & \ding{51} & \ding{51} & \ding{51} & \ding{51} & \ding{51} \\
\bottomrule
\end{tabular}
\label{tab:tabfact_capabilities}
\end{table*}


\subsection{Data Sources}
The vision behind \textit{\system} is simple. A user can drop a few \textit{grounding} data sources into SQL databases and immediately start asking the system to verify claims. We make only minimal assumptions about these data sources: they are \textit{relational}---as is often the case with official federal or state data---and we treat them as reliable and trustworthy.

\system remains completely agnostic of both the information content and the internal structure of these tables and databases. We provide no additional metadata, schema information, or prior knowledge to our multi-agent system. Instead, we assume that the \textit{grounding} data sources are entirely unknown before deployment. The agents must therefore operate in an open-ended environment, discovering and reasoning about available data as they encounter it. 

This design makes our approach highly flexible as we can quickly plug-and-play by adding or removing data sources without concern for compatibility or reconfiguration. This flexibility has been a central motivation since the inception of our system.

\subsection{Tools}

To enable this flexibility, we must address a fundamental challenge: LLMs, no matter how capable, are inherently disconnected from external data sources---they can only operate in isolation, with no way to interact with databases.  Agents bridge this gap by using \emph{tools}.  A tool acts as an interface to external capabilities,  allowing agents to interact with, perceive, and affect their environment. In general, tools can include capabilities that perform mathematical calculations, or read files from disk,  or query a database. Each agent has a fixed  collection of such tools. At runtime, the agent\footnote{More precisely, it is the LLM that makes this decision, though we often use ``agent'' and ``LLM'' interchangeably in such contexts} autonomously decides which tool to invoke, how to call it, and when to use it; the tool's output is then fed back into the reasoning loop~\cite{DBLP:conf/iclr/YaoZYDSN023}. This interactive feedback cycle between reasoning, action, and observations forms the backbone of modern agentic AI.

Building tools from scratch is challenging, because they  need to be carefully designed.  They must return well formatted values, and informative error messages, because these are fed back into the LLM.  Building a tool also requires domain expertise. For example, a simple tool that fetches schema information from a PostgreSQL database requires knowledge of relational databases, Postgres internals, and query execution.  Building  a similar tool for MySQL (another database management system), the developer has  to start over, since there are differences in the  catalog layout, the connection logic, etc.  Switching to  a different agentic framework might require rebuilding all tools from scratch.  The solution we adopted for \system was to use MCP.

\subsubsection{MCP} The Model Context Protocol (MCP), introduced by~\citet{AnthropicMCP},  standardizes how AI applications connect to different data sources, effectively eliminating the need for custom connections for each new AI model and external system allowing us to direct our energy elsewhere---away from repetitive boilerplate code.  MCP  simplifies and streamlines the tool building process, and has already been adopted by industry~\cite{PaypalMCProllout, DatabricksMCProllout, AzureMCProllout, SnowflakeMCProllout}.

\subsubsection{Toolsets} We use Google's MCP Toolbox for Databases~\cite{BlogPostGoogleMCP}, a framework that makes it effortless to organize and manage database tools. It provides built-in \textit{primitives}---actual implementations of low-level functions like executing SQL---across different database systems (e.g., PostgreSQL, MySQL), ready to use without us having to code anything.


Using these primitives as building blocks, we define higher-level tools that \textit{bind} and interact with specific databases. For example, in Figure~\ref{fig:configuration-of-tool}, we create the tools \textbf{\textcolor{yamlgreen}{\texttt{seattle\_sql}}} and \textbf{\textcolor{yamlgreen}{\texttt{portland\_sql}}}, both of which use Google’s \texttt{postgres-execute-sql} primitive to run SQL queries on the respective Postgres databases \texttt{seattle} and \texttt{portland}. We also define \textbf{\textcolor{yamlgreen}{\texttt{los\_angeles\_sql}}}, which uses the MySQL primitive \texttt{mysql-execute-sql}, to query the \texttt{los\_angeles} database.

\begin{figure}[htbp]
  \centering
\begin{minted}[gobble=4, frame=single]{yaml}
    tools:
      seattle_sql:
        kind: postgres-execute-sql # Google
        source: seattle # Postgres DB
      portland_sql:
        kind: postgres-execute-sql # Google
        source: portland # Postgres DB
      los_angeles_sql:
        kind: mysql-execute-sql # Google
        source: los_angeles # MySQL DB
      ...
\end{minted}
  \caption{A YAML fragment showing the configuration of database \textit{tools}; schema-related tools are omitted for brevity.}
  \label{fig:configuration-of-tool}
\end{figure}
Of course, there can be many such tool definitions for many different databases. Once the tools are defined, we can group them into flexible collections called \textit{toolsets}. Each agent can then simply ``subscribe'' to the toolsets it needs.

As a simple example, suppose we want an agent to investigate crime statistics across cities in the ``West Coast''---Seattle, Portland, and Los Angeles. To do that, it needs access to all three databases. All we have to do is bundle the corresponding tools from Figure~\ref{fig:configuration-of-tool} into a single \textit{toolset}, \textbf{\textcolor{yamlgreen}{\texttt{west-coast-sql}}}, and then subscribe the agent to it. It's just as easy to give the agent access to schema information: we simply subscribe it to \textbf{\textcolor{yamlgreen}{\texttt{west-coast-schema}}}. The resulting configuration is shown in Figure~\ref{fig:toolset-yaml}.


\begin{figure}[htbp]
  \centering
\begin{minted}[gobble=4, frame=single]{yaml}
    toolsets:
      west-coast-sql: 
        - seattle_sql
        - portland_sql
        - los_angeles_sql
      west-coast-schema: 
        - seattle_schema
        - portland_schema
        - los_angeles_schema
      washington-state-schema: 
        - seattle_sql
        ...
      washington-state-sql: 
        - seattle_schema
        ...
\end{minted}
  \caption{Example YAML configuration of \textit{toolsets}}
  \label{fig:toolset-yaml}
\end{figure}

In the same spirit, we might also maintain a \textbf{\textcolor{yamlgreen}{\texttt{washington-state}}} toolset, bundling together the tools for databases from the Seattle area and other cities in WA. Within each database, we can import official data from various governmental sources, which \system can then explore when verifying claims about the state---exactly as in the ongoing investigation of the City Attorney's claim from Section~\ref{sec:intro}.

If we later decide to remove access to a database (say, the Portland database), we only need to delete the corresponding tools in Figure~\ref{fig:toolset-yaml}---literally commenting out two lines of code from the configuration. Conversely, if we want to add another city into the mix, we simply append two more tools.


\subsection{Agentic System}
With the data layer now in place, we turn our attention to the core of \system: its multi-agent architecture. Connecting to databases modularly is only part of the challenge---the real difficulty lies in navigating and reasoning over them effectively. Our system tackles this through a team of three specialized expert-agents: the \textcolor{brownish}{\textit{Data Expert}}, \textcolor{blue}{\textit{Schema Expert}} and \textcolor{greenish}{\textit{SQL Expert}}. Each agent has a distinct role, specific instructions, clear output expectations, and subscribes to one of the two toolsets described earlier (\textbf{\textcolor{yamlgreen}{\texttt{sql}}} or \textbf{\textcolor{yamlgreen}{\texttt{schema}}}). 

They are coordinated by the \textcolor{redish}{\textit{Verifier}}, a higher-level agent responsible for driving the verification process and producing the final verdict on claims---along with a transparent report containing explanatory SQL queries. Importantly, the three expert-agents are designed as \textit{atomic} components: they never communicate directly with one another; they interact only with non-AI static tools exposed through their respective toolsets.

In this section, we discuss the rationale behind our design choices, the challenges we encountered, and the unique solutions that made our approach effective.

\subsubsection{Data Expert} Since the data environment is unknown, with potentially many databases and tables, we need a mechanism to rapidly survey the available landscape. This is the role of the \textcolor{brownish}{\textit{Data Expert}}, which ``subscribes'' to the \textbf{\textcolor{yamlgreen}{\texttt{schema}}} toolset. Its task is to perform a high-level scan of all accessible data sources and summarize what each source appears to contain. 

The usefulness of this step might not be immediately apparent, but it is crucial: data exploration involves numerous tool calls and exposure to large amounts of low-level information---database, table, and column names; data types, schemas, and various metadata---that must be inevitably consumed to truly understand what the data is about. The \textcolor{brownish}{\textit{Data Expert}}'s job is to ``bite the bullet'' navigating this chaos, and deliver a clean single-paragraph summary to the \textit{Verifier}. This summary enables the \textit{Verifier} to plan an effective verification strategy knowing the data sources it has in its disposal, while keeping its expensive context from being cluttered by useless details.

\subsubsection{Schema Expert} In order to write any successful query over relational databases, the first step is always to understand the schema.  In theory, relational databases should have table and column names that are unambiguous, column types should match their intended semantics (e.g., an \textit{age} column is a number and not text), and keys and foreign keys should be explicitly declared in the schema.  In practice, this is rarely the case. Corporate or institutional databases have dozens or even hundreds of tables, each with dozens of attributes, and the  table or column names are frequently non-descriptive.  For example even the relatively well organized  Crime Data for  Seattle~\cite{SeattleCrimeDashboard2025} has opaque column names like \texttt{NIBRS Group AB} or \texttt{Beat}.

This is where the \textcolor{blue}{\textit{Schema Expert}} comes into play. Its high-level role is to answer arbitrary schema-related questions about the available databases. It is equipped with the \textbf{\textcolor{yamlgreen}{\texttt{schema}}} toolset---similar to the \textit{Data Expert}---which allows it to fetch detailed schema information from the connected databases. Unlike the \textit{Data Expert}, however, it operates without guardrails and is in fact encouraged to dive deep into the structural details of the schemata. It can investigate nearly any aspect of the databases' design; from simple column names to specific constraints on those columns (e.g., foreign key relationships, nullability, and more). 

\begin{figure}[htbp]
  \centering
  \includegraphics[width=0.35\textwidth]{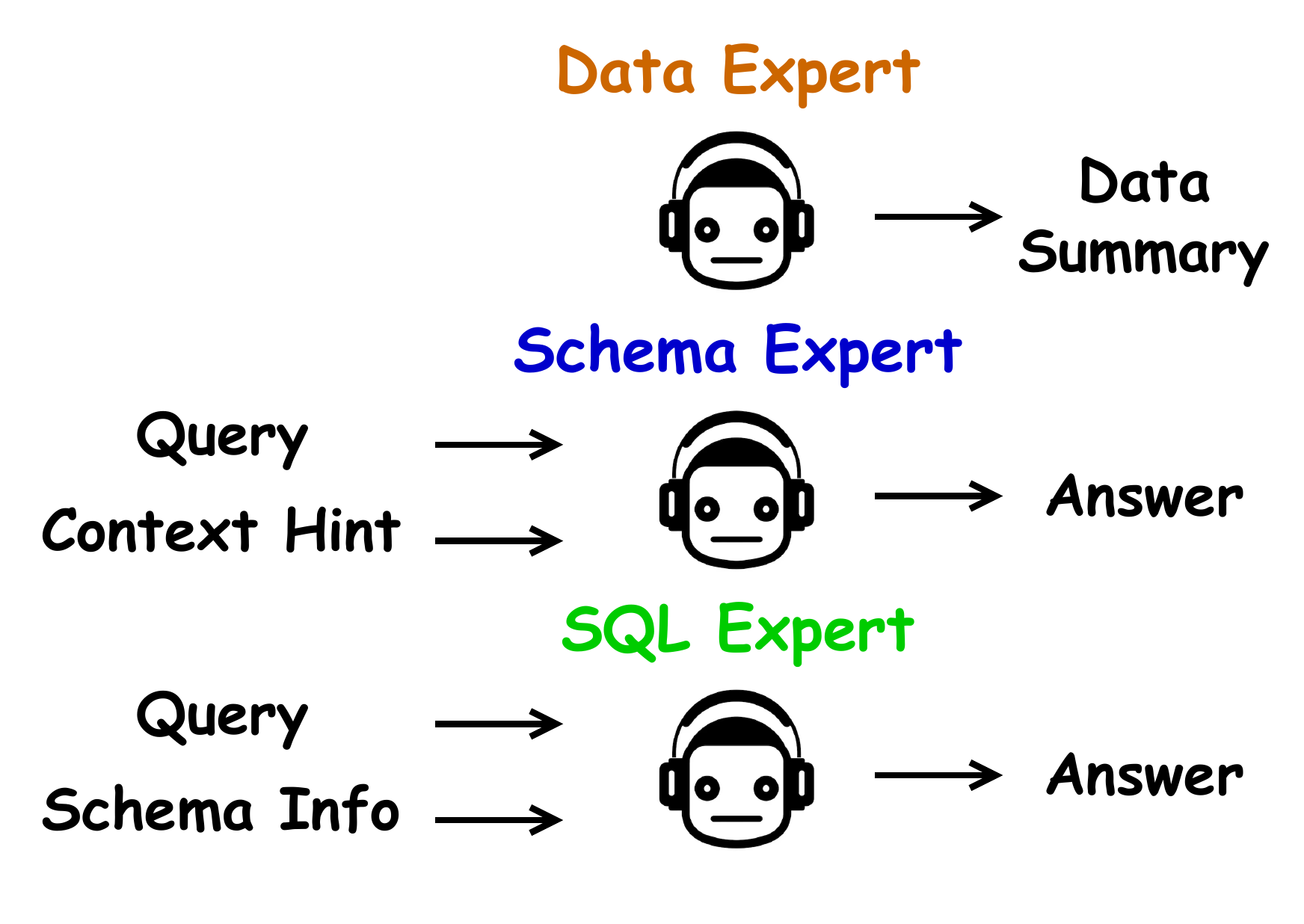}
  \caption{Inputs and outputs of the three expert-agents. The \textcolor{brownish}{\textit{Data Expert}} is invoked without input and returns a concise high-level report of what the connected databases appear to contain. The \textcolor{blue}{\textit{Schema Expert}} expects a schema-related question along with a short hint about where to look (for example, ``NYC database''), and returns a precise answer to that question. Lastly, the \textcolor{greenish}{\textit{SQL Expert}} expects a question about the data along with specific information of the schema that is relevant to the question. The process usually involves a couple of SQL queries on the databases, depending on what the agent decides. At some point, it returns a clear answer together with the concrete SQL commands that verify it.}
  \label{fig:experts_io}
\end{figure}

However, misuse of its own tools can quickly clutter the agent's memory (for example, by the misfortune of querying the schema of a messy corporate database containing several hundred-column tables). To try avoid this as much as possible, we require one additional input to the \textit{Schema Expert}. Along with the schema question, we must also provide a brief \textit{context hint}---this is a short, high-level NL cue that steers the agent toward relevant databases (Figure~\ref{fig:experts_io}).

All in all, the \textcolor{blue}{\textit{Schema Expert}} expects \textcolor{blue}{\textbf{\textbf{\circled{1}}}} a NL schema question along with \textcolor{blue}{\textbf{\circled{2}}} a \textit{context hint}, and investigates the related data sources in order to provide a crisp, precise answer in NL. The exact response and format is left to the agent and depends on the question at hand. For example, a recent query was ``\textit{List all tables related to to crime, police incidents, offense categories, or year-by-year statistics.}'' with the context hint of ``\textit{Seattle, WA}''. The resulting answer was a well-structured, markdown-formatted summary detailing the relevant tables, column names, and types.

During execution, the \textit{Verifier} frequently invokes the \textcolor{blue}{\textit{Schema Expert}} to answer different things about some database's schema (as in the example above). The responses are always clear, complete, and to the point---exactly what we want. This design ``protects'' the \textit{Verifier}'s expensive context by allowing it to access highly curated schema information without having to endure the messiness of retrieving it.


\subsubsection{SQL Expert} Once a reasonable understanding of the database's schema has been established, the next step is to interact and play around with the data itself. For relational databases, this means writing SQL. The \emph{NL to SQL} problem consists of automatically converting a natural language question over a relational database (with known schema) to an SQL query~\cite{DBLP:conf/sigmod/LiJ14,DBLP:journals/pvldb/SenLQOEDSMSS20}.  Despite lots of progress in this space and the availability of popular benchmarks---like Spider~\cite{DBLP:conf/emnlp/YuZYYWLMLYRZR18}, KaggleDBQA~\cite{DBLP:conf/acl/LeePR20} and BIRD~\cite{DBLP:conf/nips/LiHQYLLWQGHZ0LC23}---NL2SQL remains a challenging problem for real-world scenarios, because of schema complexity, query ambiguity, and semantic mismatch~\cite{DBLP:conf/cidr/FloratouPZDHTCA24}.

For example, suppose that we want to ask: \textit{Which neighborhood of Seattle recorded the most parking tickets in the second quarter of 2025?} Everyone roughly understands what this question means, yet translating it into SQL quickly reveals several challenges.  First, to compute total parking tickets, we must discover the semantics of ``quarter'' ---is it relative to the calendar year or the fiscal year? Sometimes, the best we can do is to infer it by exploring the data itself. For example, by investigating the actual data, we might find out that the table conveniently includes concatenated information that indicate the quarter (e.g., ``Q1'' to ``Q4''). Even if we are not that lucky, this process still helps us understand how dates are stored and organized. In other words, while the question is inherently ambiguous, careful inspection of the data often reveals its intended meaning.  The first stage of query formulation consists of an interactive exploration of the data.

After this exploratory stage, the next stage is  to write the query or queries that answer the question. Unfortunately, many things can still go wrong at this point. We might write a logically incorrect query (a semantic mismatch) and only realize it when no results appear---perhaps because we accidentally filtered everything out. On top of that, there are the usual syntax errors, some of which are due to the different vendor-specific versions of SQL (e.g., PostgreSQL, MySQL, Oracle, etc.) of the available databases.   

In summary, NL2SQL is not a one-shot process, but requires a continuous back-and-forth with the database through SQL queries.  In \system this is  the role of the \textcolor{greenish}{\textit{SQL Expert}}.  This expert  emulates the interactive process in order to answer questions over the available relational databases. To ensure transparency, we instruct it to always also return the concrete evidence that supports its answer. We want every result to be fully traceable back to the data through concrete SQL.  We instruct the \textcolor{greenish}{\textit{SQL Expert}}  to exclude from the answer SQL queries that are irrelevant to the final answer (e.g., exploratory or failed queries).

The \textcolor{greenish}{\textit{SQL Expert}} expects two inputs: \textcolor{greenish}{\textbf{\textbf{\circled{1}}}}~the NL query itself, and \textcolor{greenish}{\textbf{\circled{2}}} the relevant database schema information (Figure~\ref{fig:experts_io}). The latter contains all the necessary schema details the agent might need to write SQL queries within the scope of the question---such as the explicit names of the relevant databases, tables, relationships, and columns. This way, the agent can focus precisely on the relevant parts of the data, without being distracted by the rest of the environment. There might be countless other tables or databases available, but we want our \textcolor{greenish}{\textit{SQL Expert}} to enter a kind of ``tunnel-vision'' mode, concentrating solely on the question at hand and on the small relevant portion of the data landscape that contains the answer.

\subsubsection{Verifier} Finally, the role of the \textcolor{redish}{\textit{Verifier}} is to coordinate all other agents, and verify the factual claim requested by the users.  The \textcolor{redish}{\textit{Verifier}} produces two outputs: a verification verdict (one of \textit{Verified}, \textit{Partly Verified}, \textit{Partly Inaccurate}, or \textit{Inaccurate}), and an analytical report containing the SQL queries that explain and support this verdict.  This report is organized in a clear, chronological way, effectively walking the user through each stage of the verification process.  If desired, the data analyst can execute the explanation SQL queries herself, examine their outputs, modify and re-execute them, until she is completely satisfied with the veracity and generality of the claim.

To achieve this task, the \textcolor{redish}{\textit{Verifier}} interacts with all other agents in the system.  In order to ask the \textit{SQL Expert} something about the data, it will  provide it with the relevant schema information indicating where exactly to look (Figure~\ref{fig:experts_io}), obtained from the  \textit{Schema Expert}, while the latter requires information from \textit{Data Expert}.

To summarize, the \textcolor{redish}{\textit{Verifier}} begins by asking  the \textit{Data Expert} to  provide an overview of the available data sources. Using this information, the \textit{Verifier}  consults the \textit{Schema Expert} to obtain the schema of the  relevant databases,  initially at some high level of detail (for example, only the table names without their attributes).  Next, it invokes the \textit{SQL Expert} to ask a question about a specific step of the claim verification,  and obtains concrete, verifiable answers consisting of both SQL queries and their results.  Usually, more information is needed, and the \textcolor{redish}{\textit{Verifier}} repeats this in a cycle: ask the \textit{Schema Expert}  for more detail, in order to ask the  \textit{SQL Expert} new queries; if the answers remain unsatisfactory, the  \textcolor{redish}{\textit{Verifier}} may decide to  ask the \textit{Data Expert} for additional relevant data sources, and the process continues.  Eventually, the  \textcolor{redish}{\textit{Verifier}} is satisfied and writes the answer to the user.

Importantly, the \textit{Verifier} never interacts directly with the messy data sources---it leaves the ``dirty work'' to the three experts.  They are the ones who dive into the details, explore the data with whatever trials and tribulations, and handle its inevitable quirks. The \textit{Verifier} simply asks the right questions and receives informative, concise, and crisp answers in return; without ever touching the chaos underneath. Thus, its context remains light, focused, and packed only with the most useful information. This efficiency allows us to equip the \textit{Verifier} with a powerful model.

\subsection{Orchestration}
In practice, our three expert-agents are wrapped as callable functions and exposed to the \textit{Verifier} as \textit{tools}. This allows the \textit{Verifier} to invoke any of them directly, much like calling a non-AI tool like a calculator. The architecture follows an ``Agents as Tools'' pattern, where specialized agents are encapsulated as a tools with clearly defined inputs and outputs. 

For the expert-agents, we persist memory only within a single tool invocation, not across different calls. This design choice makes the agents reusable atomic components---any \textit{lead} agent, such as the \textit{Verifier}, can seamlessly employ them without inheriting messy context from previous runs. This keeps the multi-agent system simple, modular, and easy to extend.

\section{Verification Example}\label{sec:verificaton}

Now, we can turn our focus to the verification example introduced in Section~\ref{sec:intro}. As a reminder, Ann Davison, Seattle's City Attorney, made the following statement in the 2024 Annual Report:

\begin{quote}
    \textit{I am pleased to acknowledge that 2024 saw a reduction in property crime and violent crime in Seattle.}
\end{quote}

We set out to verify this claim using \system.  We have access to a PostgreSQL database with the  City of Seattle’s official crime data.  We had  downloaded this data from the \citet{SeattleCrimeDataset} in the form of a CSV file, then uploaded it into PostgreSQL.  It contains  roughly $1.5$M rows of all recorded crimes from 2008 to the present.  To verify claims like that by the city's attorney, we invoked \system with the  verbatim claim above. After a few minutes, our multi-agent system produces the following report and verdict (excerpted verbatim):

\begin{center}
\begin{minipage}{0.93\linewidth}
    \textit{\textbf{Findings}}:\vspace{-1.3mm}
    \begin{itemize}
        \item[--] \textit{Property crime increased from 40,951 (2023) to 41,220 (2024), a rise of 269 incidents ($\sim0.7\%$).}
        \item[--] \textit{Violent crime increased from 5,435 (2023) to 5,477 (2024), a rise of 42 incidents ($\sim0.8\%$).}
    \end{itemize}\vspace{-1mm}
    \textit{\textbf{Conclusion}: The claim is contradicted by the data: both property crime and violent crime were slightly higher in 2024 than in 2023 in the Seattle dataset examined.}\\
    \textit{\textbf{Verdict}: Inaccurate}
\end{minipage}
\end{center}

An important feature of \system is that it also returns the explanatory SQL query (or queries) that lead it to its verdict.  These queries can be directly inspected, executed, and reproduced by expert users.  In our example, \system generated the query shown in Figure~\ref{fig:thucy-sql-ann-davison}. In essence, the query groups crimes by year and category, and then counts the number of incidents---exactly what we would expect for this verification. The output of the query is also shown in the figure.  It was easy to run this query ourselves and confirm the correctness of \system's verdict; we show the answers in the figure.  We also checked these results on the interactive crime dashboard of the City of Seattle~\cite{SeattleCrimeDashboard2025}, and got the same results.\footnote{In the dashboard, make sure that ``all'' is selected in \textit{Precinct}.}


\begin{figure}[htbp]
  \centering
    \begin{minted}[gobble=4, frame=single]{sql}
    SELECT
     EXTRACT(YEAR FROM offense_date)::int 
     AS year,
     offense_category,
     COUNT(*) AS incident_count
    FROM public.crime_data
    WHERE offense_category IN 
        ('PROPERTY CRIME','VIOLENT CRIME')
     AND offense_date >= '2023-01-01'::date
     AND offense_date < '2025-01-01'::date
    GROUP BY 1, 2
    ORDER BY 1, 2;
    \end{minted}
    \begin{tabular}{ccc}
    \toprule
    \textbf{Year} & \textbf{Category} & \textbf{Incidents} \\
    \midrule
        2023 & Property Crime & 40,951 \\
        2023 & Violent Crime  & 5,435  \\
        2024 & Property Crime & 41,220 \\
        2024 & Violent Crime  & 5,477  \\
    \bottomrule
    \end{tabular}
  \caption{SQL query and results produced by \system when verifying the City Attorney's claim. The query groups crimes by year and category and counts total incidents.}
  \label{fig:thucy-sql-ann-davison}
\end{figure}

\section{Experiments}

In this section, we present the experimental evaluation of \system. We first describe the widely used \textit{fact verification} benchmark TabFact, followed by the baselines. Next, we outline the framework in which \system was built and the LLMs it uses. Finally, we present our findings, which show that \system decidedly surpasses the state of the art.


\subsubsection{Benchmark} We conduct experiments on TabFact~\cite{DBLP:conf/iclr/ChenWCZWLZW20}, a widely used benchmark for fact verification over Wikipedia tables. The task is to determine whether a claim holds given the evidence in a relational table. The claim is labeled ``false'' if any part of it conflicts with the data from the table. Many cases involve subtle linguistic reasoning and common sense. Following all prior work~\cite{DBLP:journals/tmlr/NguyenBSKNL25, DBLP:journals/pvldb/ZhuCXLSZSTL24, DBLP:journals/pvldb/ZhangHFCDP24}, we evaluate on the small test split of TabFact, which contains roughly 2k fact-table pairs.

\subsubsection{Baselines} We compare against recent fact-verification systems that all rely on LLMs, as these have achieved state-of-the-art performance~\cite{DBLP:journals/pvldb/ZhuCXLSZSTL24}. We do not re-implement the baselines; instead, we report the results provided in their original papers for the same task and dataset.

We compare against BINDER~\cite{DBLP:conf/iclr/ChengX0LNHXROZS23}, DATER~\cite{DBLP:conf/sigir/YeHYLHL23}, CoTable~\cite{DBLP:conf/iclr/0002ZLEP0MFSLP24}, \mbox{ReActTable}~\cite{DBLP:journals/pvldb/ZhangHFCDP24},  AutoTQA~\cite{DBLP:journals/pvldb/ZhuCXLSZSTL24}, and POS~\cite{DBLP:journals/tmlr/NguyenBSKNL25}. 

Among them, AutoTQA is particularly relevant, as it also builds a multi-agent system and is the only one in the literature to also support cross-table querying. Their agents follow a cyclic orchestration pattern---executing, critiquing, and refining plans in a loop. Our approach differs in two main ways: \circled{1} \textsc{Thucy} is agnostic to the underlying data environment, and \circled{2} it provides concrete traceable evidence alongside the answers. We also take a different stance on agent orchestration: instead of cyclic pattern, we employ decoupled, specialized expert-agents. This choice is validated by recent successful applications in industry~\cite{AnthropicMultiAgentSystem}. 

POS is also related to our work, as it focuses on \textit{interpretability}. It returns the execution plan to the user as a logical sequence of NL \textit{atomic} steps. We differ in two key ways: \circled{1} we output concrete SQL queries, eliminating any suspicion of hallucinations, since expert users can directly verify them; and \circled{2} we are not constrained to an answer coming from a single query. In contrast to POS, which assumes the final answer is produced by a \emph{single} SQL query, we allow---and in fact encourage---multi-step reasoning where potentially many arbitrary queries contribute to the final answer in different ways.

\subsubsection{Setup} We built \textsc{Thucy} using the OpenAI Agents SDK. Following our discussion in Section~\ref{sec:architecture}, we equip the \mbox{\textcolor{redish}{\textit{Verifier}}} with a highly capable model (GPT-5), since its context remains lightweight. We then experiment with the expert agents---\textcolor{brownish}{\textit{Data Expert}}, \textcolor{blue}{\textit{Schema Expert}}, and \textcolor{greenish}{\textit{SQL Expert}}---using two model variants: GPT-5-mini and GPT-4o-mini.

\subsubsection{Results} As we can see in Table~\ref{tab:tabfact_results}, \system beats the previous state of the art by 5.6 percentage points, setting a new best-known result on TabFact at $94.3\%$.  
To test the robustness of \system, we also swapped the models of our three expert agents for GPT-4o-mini, aligning them to those used in the baseline systems (e.g., we match POS). The outcome remains the same: \system outperforms the previous state of the art by 5 points in accuracy. This result is especially encouraging---it shows that \system remains effective even when the individual agents use less capable models. It also reinforces our design choice of specialized, task-specific agents, since we can confidently downgrade their models to reduce cost without sacrificing much. We believe this decomposition of the overall task into smaller, well-defined subtasks, each handled by a dedicated expert agent under a single \textit{lead} agent, plays a central role in these improvements.
%

\begin{table}[t]
\centering
\caption{Accuracy ($\uparrow$) on the small test set of the TabFact Benchmark. Some papers decided to re-run the same experiments of previous methods using newer models, so we report the new results as well. Each entry points to its source paper.}
\begin{tabular}{l c c}
\toprule
\textbf{Method} & \textbf{Model} & \textbf{Acc ($\uparrow$)} \\
\midrule
BINDER~\cite{DBLP:conf/iclr/ChengX0LNHXROZS23} & Codex & 85.1\% \\
BINDER~\cite{DBLP:journals/tmlr/NguyenBSKNL25} & GPT-4o-mini & 84.6\% \\
DATER~\cite{DBLP:conf/sigir/YeHYLHL23} & Codex & 85.6\% \\
DATER~\cite{DBLP:journals/tmlr/NguyenBSKNL25} & GPT-4o-mini & 81.0\% \\
CoTable~\cite{DBLP:conf/iclr/0002ZLEP0MFSLP24} & PaLM 2 & 86.6\% \\
CoTable~\cite{DBLP:journals/tmlr/NguyenBSKNL25} & GPT-4o-mini & 84.2\% \\
ReAcTable~\cite{DBLP:journals/pvldb/ZhuCXLSZSTL24} & GPT-4 & 83.4\% \\
ReAcTable~\cite{DBLP:journals/pvldb/ZhuCXLSZSTL24} & GPT-4-turbo & 85.0\% \\
ReAcTable~\cite{DBLP:journals/pvldb/ZhangHFCDP24} & Codex & 86.1\% \\
AutoTQA~\cite{DBLP:journals/pvldb/ZhuCXLSZSTL24} & GPT-4 & 87.4\% \\
AutoTQA~\cite{DBLP:journals/pvldb/ZhuCXLSZSTL24} & GPT-4-turbo & 88.7\% \\
POS~\cite{DBLP:journals/tmlr/NguyenBSKNL25} & GPT-4o-mini & 82.7\% \\
\midrule
\system (ours) & GPT-4o-mini & 93.7\% \\
\system (ours) & GPT-5-mini & \textbf{94.3\%} \\
\bottomrule
\end{tabular}
\label{tab:tabfact_results}
\end{table}

\section{A Journalistic Tale}\label{sec:the_story}
A very recent article by \citet{MyNorthwestClaimOct2025} claimed that violent crime in \textit{downtown} Seattle had ``plummeted'' during the summer months compared to the same period last year. The second sentence of the article reads:
\begin{quote}
    \textit{Between June and August 2025, officials reported that violent crime incidents in downtown Seattle dropped by 36\% compared to the same period in 2024.}
\end{quote}
Within just a few hours, other outlets---including \citet{Kiro7ClaimOct2025} and \citet{YahooNewsClaimOct2025}---had picked up and republished the same story, all citing the original source.

Naturally, having built \system and with the crime dataset from the \citet{SeattleCrimeDataset} already in hand, we were eager to see what it would say. We submitted the exact claim verbatim and waited a few minutes for the answer. Unlike the earlier statement by Seattle’s City Attorney (Sections~\ref{sec:intro} and~\ref{sec:verificaton}), however, this one was trickier. The term, ``downtown'', in particular, can mean different things depending on who you ask. After a few minutes, \system returned the following report (excerpted verbatim):

\begin{center}
\begin{minipage}{0.95\linewidth}
    \textit{\textbf{Conclusion}}
    \vspace{-1.3mm}
    \begin{itemize}
        \item[--] \textit{The Seattle crime data do not support the claim that “violent crime incidents in downtown Seattle dropped by 36\%” in Summer (Jun–Aug) 2025 versus the same period in 2024. Depending on how “downtown” is defined, overall violent crime either increased modestly or decreased slightly, but nowhere near 36\%.}
    \end{itemize}\vspace{-1mm}
    \textit{\textbf{Verdict}: Inaccurate}\\
    \vspace{-1mm}
    \textit{\textbf{Assumptions and notes}}
    \vspace{-0.2mm}
    \begin{itemize}
        \item[--] \textit{“Downtown Seattle” was operationalized as the neighborhoods BELLTOWN, DOWNTOWN COMMERCIAL, and PIONEER SQUARE, and also tested with CHINATOWN/INTERNATIONAL DISTRICT included.}
    \end{itemize}\vspace{-1mm}
\end{minipage}
\end{center}

\noindent What \system considered as ``Downtown Seattle'' is, in principle, reasonable. Still, before accepting the verdict, we wanted to dig deeper. The dataset includes a \texttt{neighborhood} attribute, which \system correctly\footnote{Or rather \textit{incorrectly}, as we will see in Section~\ref{sec:future}} leveraged to filter by the relevant neighborhoods. This is exactly what a data analyst would do too. But this raised an interesting question: could there exist some other combination of neighborhoods---perhaps the one implicitly used by the news articles---for which the 36\% drop actually holds?

We dug deeper into the SQL queries produced by \system. As experienced SQL users, we tweaked those queries to define ``downtown'' geographically instead: based on the distance from Seattle Central Library (which is undoubtedly \textit{downtown}). To our surprise, when we restricted to crimes only within a radius of about 0.7km, the trend of the claim begun to emerge (Table~\ref{tab:distance_table}). That only made us more determined to get to the bottom of this.

\begin{table}[ht]
\centering
\caption{Cumulative violent crime counts and percentage reduction (Jun--Aug 2024 vs. Jun--Aug 2025). Distance is counted from Seattle's Downtown Library using latitude and longitude coordinates available in the data.}
\begin{tabular}{lccc}
\toprule
\textbf{Distance} & \textbf{2024} & \textbf{2025} & \textbf{Reduction} \\
\midrule
$<0.5$km  & 30  & 37  & $-23.33\%$ \\
$<0.7$km  & 112 & 87  & $22.32$  \\
$<1.0$km  & 178 & 146 & $17.98$  \\
$<1.5$km  & 302 & 289 & $4.30$   \\
\bottomrule
\end{tabular}
\label{tab:distance_table}
\end{table}

After further investigation on the Web, we finally uncovered the source of the confusion. The original news source came from a different article, published by the \citet{OriginalClaimOct2025}, which stated:

\begin{center}
\begin{minipage}{0.93\linewidth}
\textit{Violent crime incidents in Seattle \textbf{police's M sectors} (the downtown core) declined 36\% between June–August 2025 compared to the same period in 2024.}
\end{minipage}
\end{center}

\noindent This claim is far more specific: it reveals that the 36\% refers specifically to the \emph{police's M sectors}. Admittedly, we were not familiar with this terminology. So, once again, we invoked \system with the exact wording of this claim. This time, it returned the following report:

\begin{quote}
    \textit{\textbf{Summary conclusion}}
    \vspace{-1.3mm}
    \begin{itemize}
    \item[--] \textit{Using report\_datetime (report month), violent crime incidents in Seattle Police’s sector M (downtown) fell from 105 in June–August 2024 to 67 in June–August 2025: a $-36.19$\% change, which rounds to $-36$\%. This matches the claim.}
    \end{itemize}\vspace{-1.0mm}
    \textit{\textbf{Verdict}: Verified}
\end{quote}

 \noindent With the extra \textit{M sector} information at hand, \system was able to verify it. We noticed that this time \system took a different route: it filtered the data using attributes like \texttt{sector}. We also verified \system’s findings by cross-checking the results on the interactive crime dashboard of the \citet{SeattleCrimeDashboard2025}. The numbers match perfectly.\footnote{When reproducing the results, after choosing the \textit{year} and \textit{offense category}, keep only \textit{beats} M1–M3 selected; this is sector M.}

  Even though we are not journalists, this whole process convinced us even further of the urgent need for journalistic tools that actually produce the concrete SQL evidence of their verdict. \system is one of them. It doesn't \emph{just} give a verdict---the story doesn't \emph{just} end there. It can be transformed, magnified, and turned to something greater. This is what \citet{DBLP:conf/cidr/CohenLYY11} envisioned long ago in their pioneering seminal work on \textit{computational journalism}.

\eat{
Even though we are not \textit{journalists}, this whole process made us realize how much we need journalistic tools that actually produce the concrete SQL evidence of their verdict. \system is one of them. It doesn't \emph{just} give a verdict---the story doesn't \emph{just} end there. It also shows the exact SQL queries that back it up. That is powerful because it lets anyone tweak them, question the assumptions, and take the analysis one step further. In a sense, this is what \citet{DBLP:conf/cidr/CohenLYY11} envisioned long ago in their pioneering seminal work.
}

 \section{Limitations \& Future Work}\label{sec:future}
 
  \noindent \textbf{Dirty Data.} Coming back to the previous example, when we first gave \system the ambiguous claim about \emph{downtown} Seattle, it made some assumptions about the \texttt{neighborhood}. It then filtered this attribute with the values it considered as \emph{downtown}. So far so good---but what \system missed was that this column has many missing values. In fact, roughly 50\% of them are missing. Since \system also returns the concrete SQL, we were able to spot this immediately.


  \vspace{0.3mm}
 \noindent \textbf{Assumptions \& Ambiguity.} Another direction we want to explore is controlling how much the agents rely on assumptions. Assumptions are useful as this is the only way to combat \emph{ambiguity} in both the data and user questions. However, they can also introduce subtle errors (for example, using the wrong current date). We want to experiment with ways to make these assumptions better grounded. One idea is to create another specialized expert-agent that searches the web.


  \vspace{0.3mm}
 \noindent \textbf{Quantitative Evaluation.} In the evaluation of \system, we used TabFact~\cite{DBLP:conf/iclr/ChenWCZWLZW20}. There is, however, a mismatch: we propose a system that can navigate data environments with many databases and tables, while our evaluation is conducted on a single-table benchmark. This is indeed the case, but to the best of our knowledge there is no fact-verification benchmark in the literature that focuses on large-scale cross-table or cross-database data. We believe this is an important next step for fact-verification, and we are actively working on creating one.

  \vspace{0.3mm}
 \noindent \textbf{Ablation Studies.} As we were building \system, we manually tested and refined each agent, observing both their individual behavior and their interactions within the full system. However, in this work we do not present systematic ablation studies. A careful evaluation and a systematic study of the contribution of each component---both in isolation and by removing individual agents from the system---is warranted.

 \vspace{0.3mm}
 \noindent \textbf{Stateless Expert-Agents.} Each expert agent does not preserve memory across tasks. This is a deliberate design choice, as it allows \system to operate in dynamic data environments. However, this increases cost, since agents must re-discover information across tasks. 
 

  \vspace{0.3mm}
 \noindent \textbf{Expensive.} Lastly, multi-agent systems like \system burn through tokens fast~\cite{AnthropicMultiAgentSystem}. In our case, fact-checking all 4K examples in our experiments cost about \textdollar183.9 in total. That comes out to roughly 5¢ per example. In messy real-world fact-checking scenarios like the one discussed earlier (Section~\ref{sec:the_story}), the cost rises to 20¢ per verification. 
 Still, we believe that our journalistic use case is high-stakes enough that this trade-off is worthwhile. After all, we can easily imagine journalists at the \textit{New York Times} being more than happy to spend a few dollars to have their articles \emph{stamped} by \system as \textit{verified} and fault-proof.

\section{Conclusion}\label{sec:conclusions}

We described our preliminary results for \system, the first multi-agent claim-verification system that operates over multiple relational databases and provides the concrete SQL evidence behind its verdicts. \system remains agnostic to the data environment prior to deployment and must therefore figure everything out from scratch. Our experimental results on a widely used fact-verification benchmark highlight the strength of our multi-agent design. \system improves the current state of the art in claim verification.


\section{Acknowledgments}

We thank the reviewers. This work was supported by NSF III 2507117, NSF SHF 2312195, and NSF IIS 2314527.

\bibliography{references}


\end{document}